\documentstyle[12pt,aaspp4]{article}
\def\etal{et al.}

\begin{document}

\title{$RJK$ Observations of the Optical Afterglow of GRB
991216\footnote{Based on the observations collected at the
F. L. Whipple Observatory 1.2~m telescope and the University of Hawaii
2.2~m telescope}}

\author{Peter M. Garnavich} 
\affil{University of Notre Dame, Department of Physics, 225 Nieuwland
Science Hall, Notre Dame, IN 46556}
\affil{\tt e-mail: pgarnavi@nd.edu} 

\author{Saurabh Jha, Michael A. Pahre\altaffilmark{2},
Krzysztof Z. Stanek\altaffilmark{2}, Robert P. Kirshner, Michael R. Garcia,
Andrew H. Szentgyorgyi}
\affil{Harvard-Smithsonian Center for Astrophysics, 60 Garden St.,
Cambridge, MA 02138}
\affil{\tt e-mail: sjha@cfa.harvard.edu, kstanek@cfa.harvard.edu,
mpahre@cfa.harvard.edu, mgarcia@cfa.harvard.edu,
aszentgyorgyi@cfa.harvard.edu }
\altaffiltext{2}{Hubble Fellow}

\author{John L. Tonry} 
\affil{University of Hawaii, Institute for Astronomy, 
2680 Woodlawn Dr., Honolulu, HI~96822}
\affil{\tt e-mail: jt@avidya.ifa.hawaii.edu} 

\begin{abstract}

We present near-infrared and optical observations of the afterglow to
the Gamma-Ray Burst (GRB) 991216 obtained with the F. L. Whipple
Observatory 1.2-m telescope and the University of Hawaii 2.2-m
telescope. The observations range from 15 hours to $3.8$~days after
the burst. The temporal behavior of the data is well described by a
single power-law decay $t^{-1.36\pm 0.04}$, independent of wavelength.

The optical spectral energy distribution, corrected for significant
Galactic reddening of $E(B-V)=0.626$, is well fitted by a single
power-law with $\nu^{-0.58\pm 0.08}$. Combining the IR/optical
observations with a Chandra X-ray measurement gives a spectral
index of $-0.8\pm 0.1$ in the synchrotron cooling regime.
A comparison between the
spectral and temporal power-law indices suggest that a jet is a better
match to the observations than a simple spherical shock.

\end{abstract}


\section{INTRODUCTION}

The BeppoSAX (Boella et al.~1997) and RXTE (Levine \etal\ 1996) satellites
have brought a new dimension to gamma-ray burst (GRB) research, by
providing rapid localizations of several bursts per year. This
has allowed many GRBs to be followed up at other wavelengths,
ranging from the X-ray (Costa et al.~1997) and optical (van Paradijs et
al.~1997) to the radio (Frail et al.~1997).  Precise positions have also
allowed redshifts to be measured for a number of GRBs (e.g. GRB
970508: Metzger et al.~1997), providing definitive proof of their
cosmological origin.

The extremely bright gamma-ray burst GRB 991216 was detected by BATSE
(Kippen, Preece, \&  Giblin~1999) on December 16.671544 UT, with its peak flux
(fluence) ranking it as the 2nd (13th) of all BATSE bursts detected so
far. The RXTE PCA search for the X-ray afterglow of GRB 991216 started
about four hours after the burst (Takeshima et al.~1999) and detected
a strong, decaying X-ray afterglow, providing much improved burst
position. It should be noted that the X-ray afterglow of GRB 991216
was also detected by much less sensitive RXTE ASM instrument as early
as one hour after the burst (Corbet \&  Smith~1999), providing a
measurement of the X-ray afterglow at times which have previously not
been studied. In addition, observations of GRB 991216
by the Chandra Observatory resulted in
the first arcsecond position determination for an X-ray afterglow 
(Piro et al. 1999).

The optical afterglow of GRB 991216 was identified by Uglesich et
al.~(1999) with data taken about $12$~hours (December 17.142 and
17.372 UT) after the burst, using the MDM 1.3-m telescope. It was
recognized as a bright variable object ($R\approx18.8$ at
Dec. 17.142), not present in the digitized POSS II plate, declining
with a temporal decay index of $\approx -1.4$.  Numerous independent
confirming observations of the fading optical transient (OT) have
followed, starting with Henden et al.~(1999) and Jha et al.~(1999).
Near-infrared observations were also reported by Vreeswijk \etal\
(1999a) and Garnavich \etal\ (1999b).

Absorption lines at $z=1.02$ seen in the optical spectrum of
GRB~991216 taken with the VLT-UT1 8-m telescope by Vreeswijk et
al.~(1999b) provide a lower limit to the redshift of the GRB source.
Given the gamma-ray fluence (Kippen 1999), the isotropic
energy from the burst was more than 8$\times 10^{53}$~ergs
($H_o=65$ km$\;$s$^{-1}\;$Mpc$^{-1}$, $\Omega_m=0.3$, $\Omega_\Lambda=0.7$),
or nearly half a Solar rest mass radiated away in under 10 seconds. 
This exceedingly large energy requirement
can be reduced if the burst emission is beamed.
To date, evidence for jets has been found in only a handful of
GRB afterglows (Sari, Piran \& Halpern 1999; Kulkarni et al. 1999;
Stanek et al. 1999) and it remains to be shown whether anisotropy
is ubiquitous.

We present optical and near-IR photometry of GRB~991216 from
observations obtained at the Hawaii 88-inch and the Fred L. Whipple
1.2m telescopes.  We describe the data and the reduction procedure in
Section 2.  In Section 3 we discuss the multiband temporal behavior of
the GRB OT.  In Section 4 we describe the broad-band spectral
properties of the afterglow deduced from our IR/optical data.

\section{OBSERVATIONS}

\begin{figure}[t]
\plotfiddle{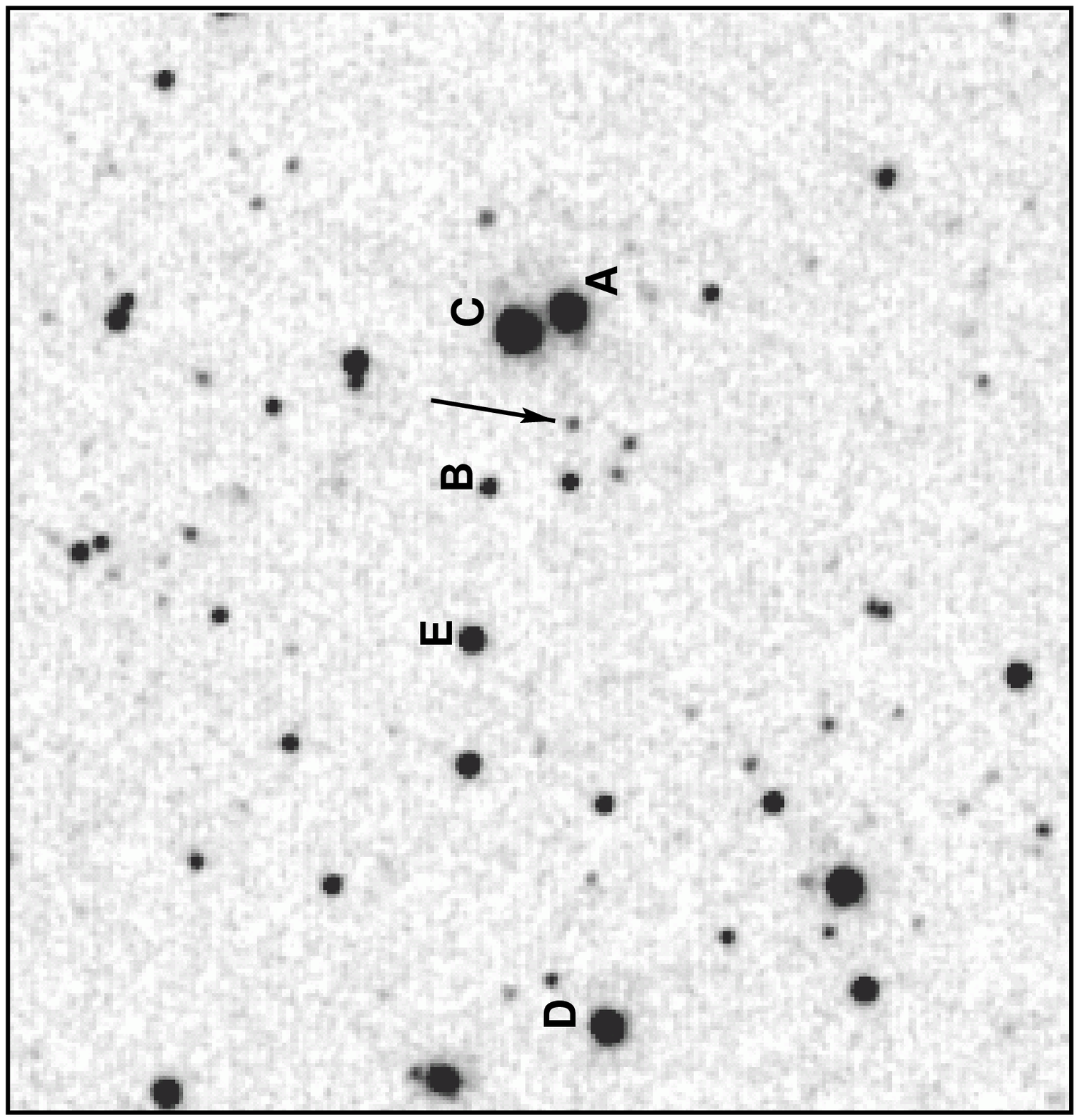}{10.0cm}{-90}{60}{60}{-30}{350}
{{\bf Fig 1. }Finding chart for the field of GRB 991216 taken in the
$J$-band.  The optical transient is indicated with an arrow and stars
calibrated as secondary standards are shown using the convention of
Jha \etal\ (1999).  North is up, east to the left and the field is
approximately 3$'$ on a side.}
\label{fig:field}
\end{figure}

The near-infrared data were obtained with the Fred L. Whipple
observatory 1.2-meter telescope on four consecutive nights beginning
1999 Dec. 17.22 (UT).  Images were taken with the ``STELIRCAM''
two-channel IR camera which utilizes two $256^2$ pixel HgCdTe arrays.
A dichroic mirror splits the beam at $\lambda \approx 1.8 \mu$m
allowing simultaneous observations in two filters. The GRB afterglow
was observed in J and K filters manufactured by Barr. The camera has
three sets of re-imaging optics and we employed the 5$'$ field-of-view
with a 1.2$"$ per pixel scale.

We immediately began imaging the RXTE localization error box after
being notified of a bright GRB detected by BATSE through the GCN
Circulars.  A 3$\times$3 mapping (15$'$ field) around the
initial RXTE position was
performed with two 60 second exposures taken at each pointing. A
$J=17$ mag object that did not appear on the digitized sky survey was
tentatively identified as the afterglow candidate (Garnavich et al. 1999a),
however it was
pointed out by Diercks \etal\ (1999a) that the star appeared on the
POSS-II N emulsion photographs and was likely to be a very red
star. RXTE revised its error box 8$'$ northward from the original
position and a new mapping was begun. Uglesich \etal\ (1999) then
identified the true afterglow near the revised position soon after
observations at FLWO were terminated. Fortunately the original mapping
and the mapping centered on the revised RXTE error box included the
object in several of the images. In subsequent nights GRB 991216 was
observed in $J$ and $K$ with 9$\times$60 second exposure sets. An
extensive number of Persson \etal\ (1998) standards were observed on
Dec. 18 (UT) and used to calibrate stars in the GRB field (Table~1; Figure~1). 
Our $J$ and $K$ calibrations are in good agreement with that of Henden, Guetter, \&
Vrba (2000), but our GRB magnitudes are 20\% to 30\% fainter in $J$
and brighter in $K$ than the
Vreeswijk \etal\ (1999a) infrared photometry. 

After the optical counterpart was identified, a single exposure of the
field was obtained with the University of Hawaii 88-inch telescope. By
then, the target was well past the meridian and the data suffered from
a high airmass. On Dec. 18 (UT) the field was observed at two epochs
and Landolt standards (Landolt 1992) imaged to calibrate the data. Since GRB~991216
was only observed in the $R$ filter, no color correction was possible
and we estimate the uncertainty from unknown color term as 5\% .  Our
$R$-band calibration of stars A and B are in good agreement with that
found by Dolan \etal\ (1999). Our final $R$-band magnitudes are on average 0.08~mag
fainter than the preliminary magnitudes given in Jha et al. (1999) and
Garnavich et al. (1999b).

From our optical imaging we find a position for the transient of
$\alpha =05^h09^m31.29^s$ $\delta =+11^\circ 17'07.3"$ (J2000)
with an accuracy of $\pm 0.\arcsec
2$ based on positions from the USNO A2.0 catalog (Monet \etal\ 1996).

\section{THE TEMPORAL BEHAVIOR}

\begin{figure}[t]
\plotfiddle{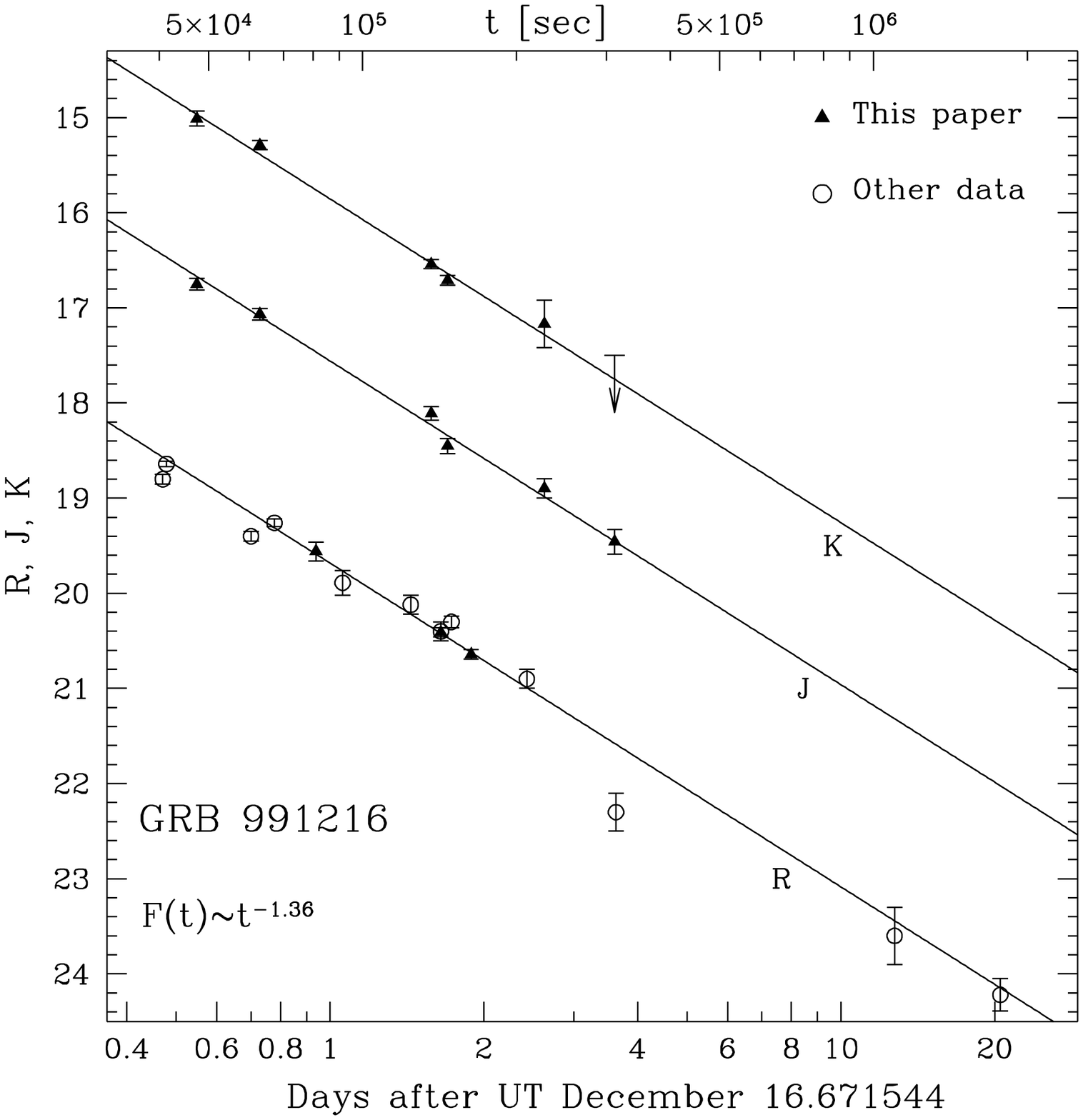}{11cm}{0}{60}{60}{40}{-80}
{{\bf Fig 2. }$RJK$ light curves of GRB 991216. Our data is shown with
solid points and other observations published in the GCN Circulars are
shown as open points. Also shown is the single power-law fit obtained
by combining all of our data.}
\label{fig:time}
\end{figure}

Figure~2 shows the $RJK$ light curves of GRB991216. Additional
$R$-band points obtained from GCN Circulars (Uglesich et al. 1999;
Dolan et al. 1999; Vreeswijk et al. 1999a; Diercks et al. 1999b; Jensen
et al. 1999; Leibowitz et al. 1999; Mattox 1999) are also plotted, but
the comparison stars used in their calibration are sometimes not known
and these points are used here only to confirm the trends seen in our
data. Late-time observations by Djorgovski et al. (1999) and Schaefer
(2000) use the Dolan \etal\ (1999) or Jha \etal\ (1999)
calibrations and are consistent with our points. The light curves appear
to follow a single power-law between 0.5 days and four days after the
burst. So not to confuse the temporal and spectral variations,
we will use the convention that $F_\nu\propto t^{-\alpha}\nu^{-\beta}$.

A single power-law was fitted to our data points by allowing the magnitude
shift between $J$ and $K$ and the shift between $J$ and $R$ be free
parameters. The result is shown as the solid lines in Figure~2 and
provides an index of $\alpha =1.36\pm 0.04$ ($1\sigma$).  Fitting the
individual bands gives indices of $\alpha =1.44\pm 0.06$ for $K$, $1.31\pm
0.06$ for $J$ and $1.42\pm 0.16$ for $R$.  Combining our three
$R$-band observations with six observations from the GCN Circulars
obtained within four days of the burst gives a power-law index of
$\alpha =1.30\pm 0.05$, somewhat steeper than the decay rate found by
Sagar et al. (2000) from the raw GCN $R$-band magnitudes. 
Clearly, a power-law index of $\alpha =1.36$ is a good fit
to all three bands given the estimated errors.  Extrapolating the $R$
fit to the late-time observations by Schaefer (2000) and Djorgovski
\etal\ (1999) shows that the single power-law is consistent with the
data out to 20 days after the burst.  The $R$-band point by Mattox
(1999) appears significantly below the trend. Our $J$-band photometry,
obtained near that time, shows no deviation from the fit, however, we
can not rule out a change in slope beginning four days after the burst
and then a recovery at late-times due to a possible underlying
supernova or host galaxy.

\section{REDDENING AND BROAD-BAND SPECTRAL ENERGY DISTRIBUTION}

\begin{figure}[h]
\plotfiddle{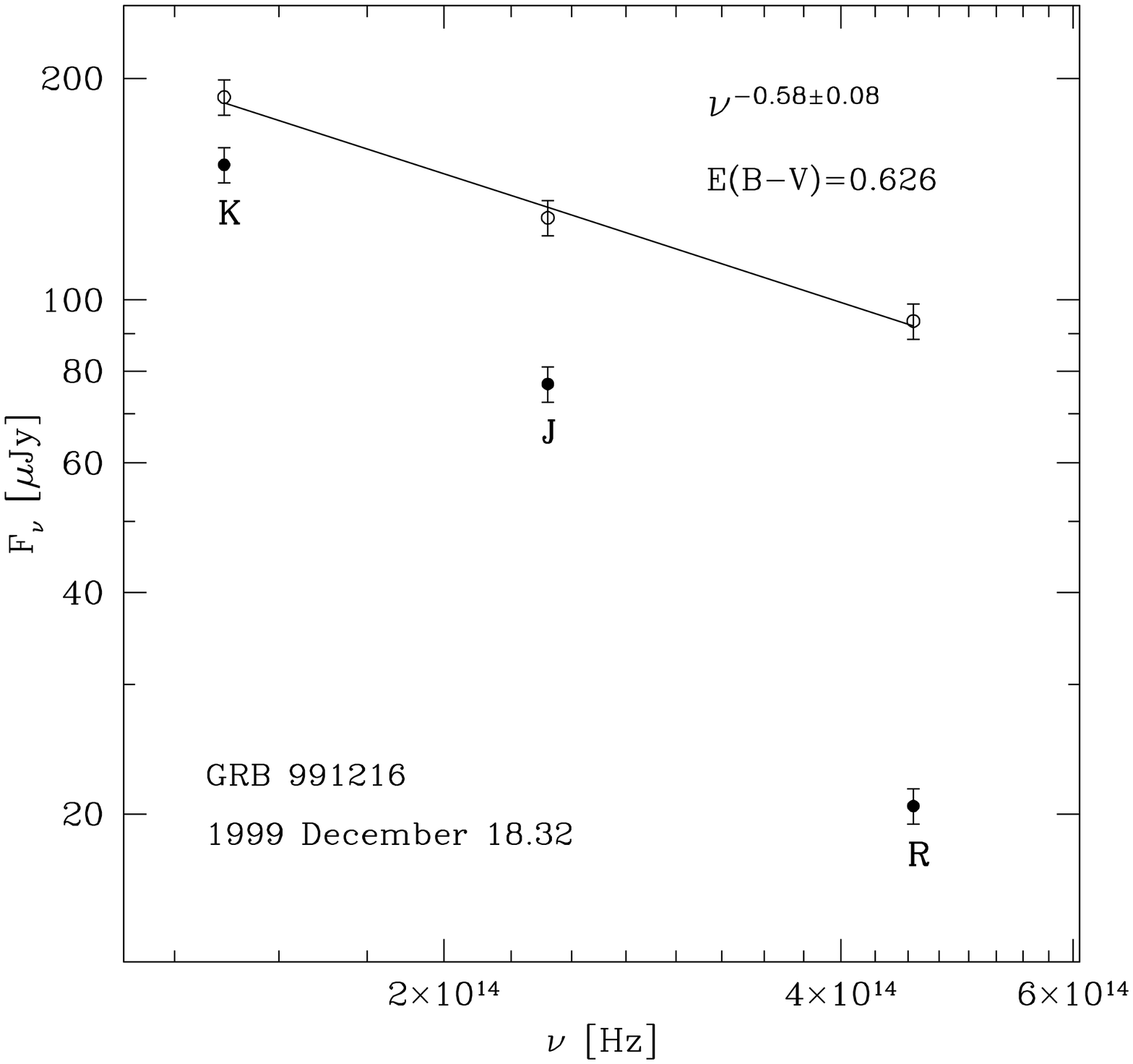}{11cm}{0}{60}{60}{40}{-80}
{{\bf Fig 3. }Synthetic spectrum of GRB 991216 $40$ hours after the
burst. The points at the top show the data corrected for a
reddening of E($B-V$)=0.626 mag.}
\label{fig:spectrum}
\end{figure}

The GRB 991216 is located at Galactic coordinates of
$l=190\arcdeg\!\!.44, b=-16\arcdeg\!\!.63$. To remove the effects of
the Galactic interstellar extinction we used the reddening map of
Schlegel, Finkbeiner \& Davis (1998, hereafter: SFD). The expected
Galactic reddening towards the burst is substantial, $E(B-V)=0.626$ mag.
We use $R_V=3.1$ and the standard reddening curve of Cardelli, Clayton \& Mathis
(1989), as tabulated by SFD (their Table~6), to correct our optical
and IR data. As discussed by Stanek et al.~(1999), there is some
indication that the SFD map overestimates the $E(B-V)$ values by a
factor of 1.3-1.5 close to the Galactic plane ($|b|<5^\circ$) (Stanek
1998) and in high extinction ($A_V > 0.5\;$mag) regions (Arce \&
Goodman 1999). It is not clear at all that such a correction should be
applied to the SFD $E(B-V)$ value for the GRB 991216, but it would
reduce this value to about $E(B-V)=0.46$.

We synthesize the $RJK$ spectrum from our data by interpolating the
magnitudes to a common time.  As discussed in the previous section,
the colors of the GRB 991216 counterpart do not show significant
variation.  We therefore select an epoch of Dec. 18.32 UT ($40$ hours
after the burst) for the color analysis, which is near the time when
simultaneous $RJK$ data were taken.

We convert the $RJK$ magnitudes to fluxes using the effective
wavelengths and normalizations of Fukugita, Shimasaku \& Ichikawa
(1995) for the optical and M\'egessier (1995) for the IR. These
conversions are accurate to about 5\%, which increases the error-bars
correspondingly. Note that while the error in the $E(B-V)$ reddening
value has not been applied to the error-bars of individual points, we
include it in the error budget of the fitted slope. The results are
plotted in Figure~3 for both the observed and the
dereddened magnitudes.  The corrected spectrum is well fitted by a
single power-law with $\beta =0.58\pm 0.08$.  If we use the lower
value of $E(B-V)=0.46$, as discussed above, the corresponding number
is $\beta =0.87\pm 0.08$. We have assumed that there is no
extinction within the host galaxy, but any reddening from the host
will make the intrinsic spectrum more flat and reduce the derived
value of $\beta$.

A radio observation (Taylor \& Berger 1999; Frail et al. 2000) found
the afterglow at
8.5 GHz to be 960$\pm 67$ $\mu$Jy on 1999 Dec. 18.16 . Adjusting our
$K$-band flux to this date, we find a power-law index between the
radio and near-IR to be $-0.15$, much more shallow than the index
between the near-IR and the optical. The Chandra X-ray observatory
also observed the GRB on Dec 18.2 (UT) (Piro \etal\ 1999) and
converting to a flux density we find a power-law index
between the IR and X-ray points of $\beta =0.8\pm 0.1$, slightly steeper
than a simple extrapolation from the IR/optical data. We note
that this slope is close to what is found for the IR/optical
data if the extinction is set to the lower value of the range discussed
above.
Figure~4 shows the overall spectrum from the radio to the X-rays and
evidence for a spectra break at frequencies less than the $K$-band.

\begin{figure}[h]
\plotfiddle{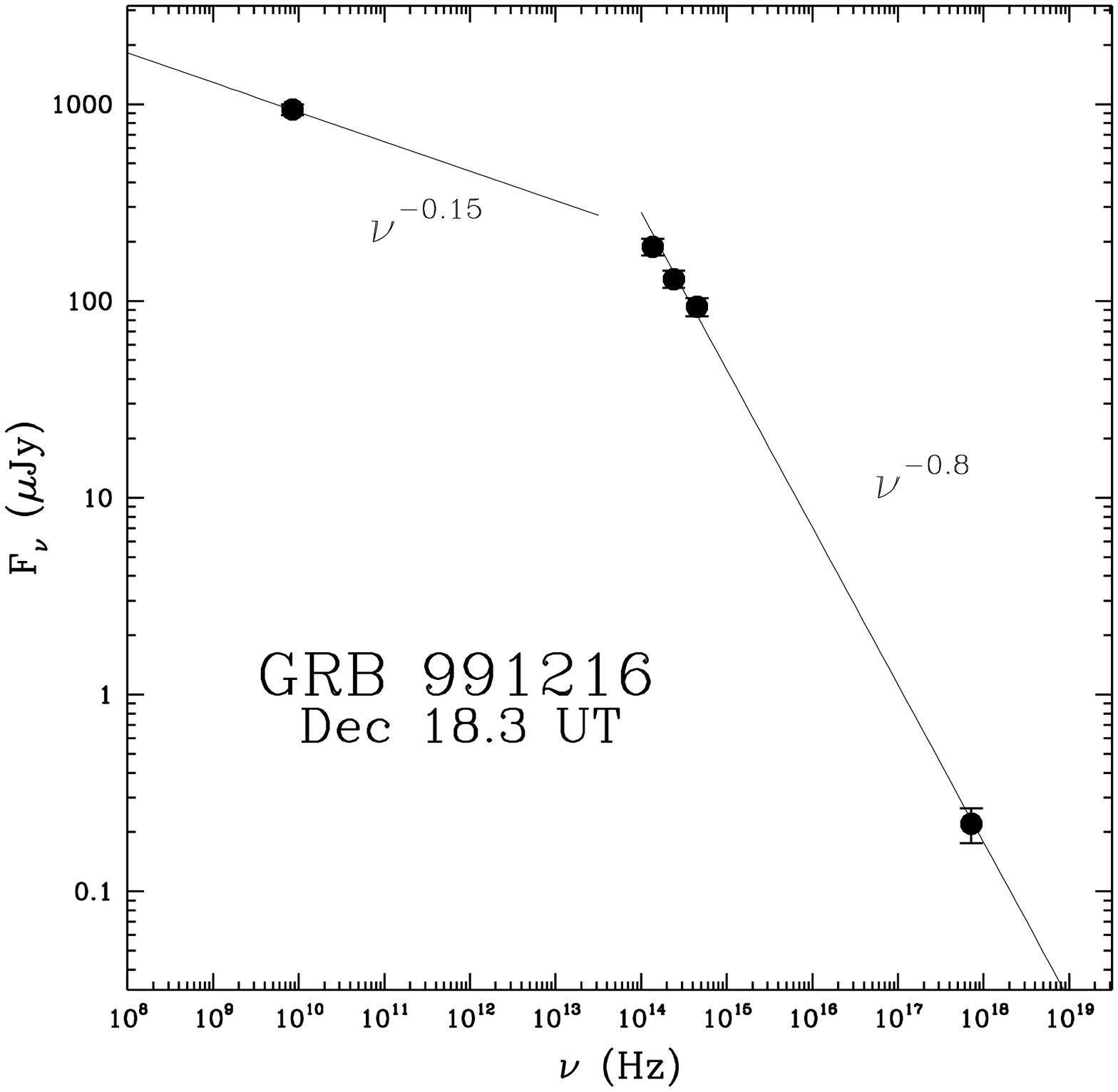}{11cm}{0}{60}{60}{40}{-80}
{{\bf Fig 4. }The spectrum of GRB~991216 over 10 orders of magnitude in
frequency on Dec. 18.3 (UT). The radio data are from Taylor \& Berger (1999)
and Frail et al. (2000). The X-ray point is derived from Piro et al. (1999).}
\label{fig:fullspectrum}
\end{figure}

The GRB afterglow model described by Sari, Piran, \& Halpern (1999)
can be used to compare the spectral and temporal power-law indices
observed. 
The IR/optical region is within the cooling regime (Figure~4), so the
observed spectral slope of $\beta =0.6$ ($0.8$ for IR to X-ray) gives an estimate of the
electron distribution index of $p=1.2$ ($1.6$).  For a spherical shock, we
then expect a temporal index of $\alpha =(3\beta -1)/2=0.4$ ($0.7$) which is
much more shallow than the observed index. For a jet, however, the
expected light curve index is $\alpha =2\beta =1.2$ ($1.6$), close to
the observed value of 1.4.  At frequencies less than the cooling
break, we expect a spectral index of 0.1 based on the IR/optical
slope. This is similar to the observed index of 0.15, but it should be
noted that other spectral breaks may be present between the radio and
IR points.

\section{CONCLUSIONS}

We present well-calibrated $RJK$ observations of the GRB~991216. 
Our data indicates
that the decay of the optical afterglow is well
represented by a single power-law with index $\alpha =1.36\pm 0.04$ from 0.5
days to four days after the burst. Combining published late-time
$R$-band observations with our data suggests a single power-law is a
good fit out to 20 days after the burst.

The optical spectral energy distribution, corrected for significant
Galactic reddening, is well fitted by a single power-law with
an index of $\beta =0.58\pm 0.08$. However, when the possible systematic error in
the SFD extinction map is considered, the index may be somewhat
steeper ($\beta =0.87\pm 0.08$). A Chandra X-ray observation obtained near
the time of our photometry provides a spectral index between the
near IR and X-rays of $\beta =0.8\pm 0.1$.

A comparison between the spectral and temporal power-law indices
suggest that the GRB is not consistent with a simple spherical shock
model. The IR/optical light curve and colors are better matched by a
shock produced from a collimated jet.

\acknowledgments{S. Barthelmy, the organizer of the GRB Coordinates
Network (GCN), is recognized for his extremely useful effort.  Support
for M. A. P. (HF-01099.01-97A) and K. Z. S. (HF-01124.01-99A) was provided
by NASA through Hubble Fellowship grants from the Space Telescope Science
Institute, which is operated by the Association of Universities for
Research in Astronomy, Inc., under NASA contract NAS5-26555. RPK  and
SJ acknowledge NSF support through AST98-19825 and a
NSF Graduate Research Fellowship.}


\tablenum{1}
\begin{planotable}{ccccc}
\tablewidth{40pc}
\tablecaption{\sc Secondary Standards near GRB~991216}
\tablehead{ \colhead{Star} & \colhead{R.A (J2000) DEC} & \colhead{R} & 
\colhead{J} & \colhead{K}  }
\startdata  
   A & 05:09:29.80 11:17:08.4 & 15.38 (02) & 13.34 (03) &  12.52 (03) \\
   B & 05:09:32.14 11:17:23.6 & 19.53 (05) & 17.07 (06) &  16.15 (08) \\
   C & 05:09:30.07 11:17:18.3 & \nodata    & 12.60 (03) &  11.80 (03) \\
   D & 05:09:39.29 11:16:59.4 & 15.21 (02) & 13.73 (03) &  13.23 (05) \\
   E & 05:09:34.16 11:17:26.5 & 18.43 (04) & 15.19 (05) &  14.14 (06) \\
\enddata
\label{table:stars}
\end{planotable}


\tablenum{2}
\begin{planotable}{ccccc}
\tablewidth{30pc}
\tablecaption{\sc GRB 991216 $J$ and $K$-Band Lightcurves}
\tablehead{ \colhead{UT} & \colhead{Hours after} & \colhead{$J$} & \colhead{$K$} & Exposure \\  
\colhead{1999 Dec.} & \colhead{the burst} & &  & (min) }
\startdata  
   17.22 & 13.2 & 16.75 (06) & 15.01 (08) & 2 \\
   17.40 & 17.5 & 17.07 (06) & 15.29 (05) & 8 \\
   18.25 & 37.9 & 18.11 (07) & 16.54 (07) & 9 \\
   18.37 & 40.8 & 18.45 (08) & 16.71 (08) & 9 \\
   19.30 & 63.1 & 18.90 (10) & 17.17 (25) & 18 \\
   20.28 & 86.6 & 19.46 (13) &  $>17.5$   & 25 \\
\enddata
\label{table:JK}
\end{planotable}


\tablenum{3}
\begin{planotable}{ccc}
\tablewidth{20pc}
\tablecaption{\sc GRB 991216 $R$-Band Lightcurve}
\tablehead{ \colhead{UT} & \colhead{Hours after} & \colhead{$R$} \\  
\colhead{1999 Dec.} & \colhead{the burst} & }
\startdata  
   17.61 & 22.5 &  19.56 (10)  \\
   18.32 & 39.6 &  20.41 (05)  \\
   18.56 & 45.3 &  20.64 (05)  \\
\enddata
\label{table:R}
\end{planotable}

\end{document}